\documentclass[runningheads]{llncs}
\usepackage{graphicx}
\usepackage{comment}
\usepackage[dvipsnames]{xcolor}
\usepackage[nolist]{acronym}
\usepackage{enumitem}
\usepackage{subcaption}
\usepackage{setspace}
\newcommand{\et}[1]{#1~et~al.}

\begin{document}
%
\title{Creating a Robot Coach for Mindfulness and Wellbeing: A Longitudinal Study}


%
\titlerunning{Creating a Robot Coach for Mindfulness and Wellbeing}
%

\author{Indu P. Bodala \and Nikhil Churamani \and Hatice Gunes}

\authorrunning{Bodala et al.}
%
\institute{Department of Computer Science and Technology\\
University of Cambridge, Cambridge, UK\\
\email{\{ipb29, nikhil.churamani, hatice.gunes\}@cl.cam.ac.uk}}

\maketitle              
\begin{abstract}
Social robots are starting to become incorporated into daily lives by assisting in the promotion of physical and mental wellbeing. This paper investigates the use of social robots for delivering mindfulness sessions. We created a teleoperated robotic platform that enables an experienced human coach to conduct the sessions in a virtual manner by replicating upper body and head pose in real time. The coach is also able to view the world from the robot’s perspective and make a conversation with participants by talking and listening through the robot. We studied how participants interacted with a teleoperated robot mindfulness coach over a period of 5 weeks and compared with the interactions another group of participants had with a human coach. The mindfulness sessions delivered by both types of coaching invoked positive responses from the participants for all the sessions. We found that the participants rated the interactions with human coach consistently high in all aspects. However, there is a longitudinal change in the ratings for the interaction with the teleoperated robot for the aspects of motion and conversation. We also found that the participants' personality traits -- conscientiousness and neuroticism influenced the perceptions of the robot coach. 

\end{abstract}

\section{Introduction}



The need for mental wellbeing interventions for general populations is ever increasing \cite{barnes2017examination,sutcliffe2016mindfulness}. Researchers have investigated the efficacy of mindfulness training in alleviating anxiety and depression, improving resilience in students towards stress, fostering emotion regulation strategies and lowering depressive moods~\cite{galante2018mindfulness,wimmer2019improving}. Particularly, the current COVID-19 climate has caused many people to experience a sense of vulnerability, loss of control, and challenging emotions such as fear and grief~\cite{ho2020mental}. Mindfulness practice has been recommended as one potential solution to this situation by encouraging to pay attention to present moment experiences with curiosity and compassion. Mindfulness is also gaining prominence as a wellbeing approach and a life-skill that can be attained with regular practice rather than just being a treatment or therapy.

Despite the benefits, learning and practising mindfulness can be challenging and inaccessible due to lack of  trained teachers and training programs, misconceptions about the methods involved and difficulties with establishing a guided regular practice.  Virtual conversational agents, in the form of mobile applications and chat-bots, have been developed to improve the accessibility of mindfulness training~\cite{hudlicka2017enhancing,sliwinski2017review}. However, majority of these agents are not interactive in nature and when they are, they rely on text-based and non-adaptive communication. Social robots with multi-modal interaction capabilities of speech, gestures and vision, constitute a promising solution.

In this study, we investigate the use of robots to deliver mindfulness sessions. For this purpose, we conducted a 5-week mindfulness study where two independent groups received mindfulness sessions, once a week, from an experienced human coach and the Pepper robot\footnote{\scriptsize{https://www.softbankrobotics.com/emea/en/pepper}} teleoperated by the human coach respectively. Interaction sessions with the teleoperated robot were made as natural as possible by replicating the upper-body and head movements of the human coach on the robot, in real-time. The robot also maintained a coherent conversation with the participants where the human coach teleoperating it could speak and listen to the participants in real-time while looking at them. In this paper, we present the results of our experiments, comparing the changes in perception of the participants, over time, towards the human and the teleoperated robot coach. 


\section{Related Work}

\subsection{Social Robots for Mental Wellbeing}
The design and use of social robots to support wellbeing is an active field of investigation~\cite{riek2016robotics}. Interactions with the Paro robot~\cite{wada2007living,turkle2006relational}, showed positive effects on wellbeing, social facilitation as well as coping with physical and cognitive impairments in elderly populations. Additionally, children with \ac{ASD} also benefited from similar robot-assisted therapies that enable embodied interactions, increasing engagement and attention and decreasing social anxiety~\cite{dautenhahn2004towards}. 
Robot coaches were also employed to instil behavioural changes in people while dieting where the robot can help individuals keep a track of their weight loss~\cite{kidd2008robots}.

Despite positive results, the field of \ac{SAR} still faces a major challenge with respect to acceptance by various stakeholders. The primary concern comes from the psychologists' lack of confidence in using robots in their practice~\cite{conti2019future,diehl2012clinical}. Autonomous robots also raise ethical concerns and may not be accepted by the general public~\cite{coeckelbergh2016survey}. To address some of these concerns, \et{Winkle}~\cite{winkle2019mutual} proposed the mutual shaping approach to the design of \ac{SAR} and measured significant shift in participants' acceptance of robots pre- and post-study. \et{Cao}~\cite{cao2019robot} proposed a supervised autonomous therapy system that enables therapists to take full control of the therapeutic environment, using the robot as an effective tool, rather than being replaced by it. Other aspects of the robots such as their appearance, embodiment, personality, engagement and adaptation to interactions were also studied to make the design of social robots more acceptable~\cite{de2013exploring}.

\subsection{Longitudinal Studies and Data Analysis}

Longitudinal studies are extremely useful to investigate changes in user behaviour and experiences over time. The motivation behind the investigation of long-term effects in human-robot interaction is that current robots and virtual agents lack social capabilities to engage users over extended periods of time. Some of the early long-term studies showed that the novelty effect quickly wears off and people lose interest and change their attitudes towards robots~\cite{gockley2005designing}. This may result in a decreased use of robots and adherence to the interventions being delivered~\cite{wang2010rome,winkle2018social}. \et{Leite}~\cite{leite2013social} presented a detailed survey on the use of social robots for long-term interactions and discussed future directions and design considerations to promote their long-term use. \et{De Graaf}~\cite{de2017they} investigated reasons for the refusal and abandonment of social robots through questionnaires and interviews. Their findings suggest that a key challenge in designing robots for long-term use is to make them easy to use in the short-term as well as functionally-relevant in the long-term. Longitudinal studies also provide us an opportunity to study and design advanced strategies for robots' continuity and incremental behaviours, affective interactions, and memory and adaptation to create a framework that engages users over continued interactions.

The data collected over a longitudinal interaction study is characterized by inter- and intra-individual variability in the overall values as well as the patterns of change. Hence, \ac{GLM} methods such as repeated measures \acs{ANOVA} may not accurately explain longitudinal changes as they only look at the changes in group means and variances~\cite{weinfurt2000repeated}. Moreover, it is important to understand that the total variance cannot be interpreted as random error in longitudinal data. Growth modelling can address these issues as we can look into how individuals change over time and whether there are any differences in their change patterns~\cite{ployhart2010longitudinal}. \ac{RCM} is such an approach that can model group and individual level variability~\cite{bliese2002growth}. Univariate longitudinal models can also be extended to analyze multivariate and more complex models.

\section{Methods}

\subsection{Experiment Design}
Our objective is to study the differences in perceptions and the overall effectiveness of mindfulness practice experienced by participants when delivered by a human coach versus a teleoperated robot coach. We conducted a between-subjects study where participants were randomly assigned to two groups - one group taught by an experienced \ac{HC} and the other taught by the Pepper robot teleoperated by the same human coach (\acs{RC}) (see Section~\ref{teleop}).  

The study consisted of $5$ mindfulness sessions, with one session administered each week for around $40$ minutes. Each session focused on a specific topic, and the $5$ sessions together were designed to provide an introduction to mindfulness techniques and suggest how to integrate mindfulness into daily life. These sessions were structured along several types of pedagogical strategies necessary for effective coaching of mindfulness meditation: didactic, to convey the conceptual basis of mindfulness; experiential learning, via guided meditations; and dyadic interactions designed to maintain participant engagement via peer and coach dialogue customized for each session with the use of external props.

We recruited staff and students across the University of Cambridge, splitting them in two groups; \ac{HC}, consisting of $2$ males and $7$ females guided by the human coach and \acs{RC}, consisting of $6$ males and $3$ females guided by the teleoperated robot coach. They were offered an incentive in the form of Amazon vouchers upon attending all the sessions. $8$ out of the $9$ participants in the \ac{HC} group, and $5$ out of the $9$ participants in \acs{RC} group attended all the $5$ sessions and the others missed one session each at random.

The experiment design was reviewed and approved by the Ethics Review Board of the Department of Computer Science and Technology, University of Cambridge. All the participants provided informed consent for the recording of audio and video data during the session for further analysis. Participants were advised that they should not be undertaking other professional mental health related treatments or medication if they plan to take part in this study. We also asked the participants to fill PHQ9~\cite{kroenke2001phq} and GAD7~\cite{spitzer2006brief} questionnaires to assess depression and anxiety scores, respectively. We reviewed the scores to make sure none of the participants were experiencing high anxiety or depression levels. 

The participants filled a $20-$item personality questionnaire at the beginning of the study~\cite{topolewska2014short}. After every session, the participants filled a `Session Experience' questionnaire about their human or teleoperated robot coach. For the \acs{RC} sessions, the questionnaire is adapted from a combination of the Godspeed~\cite{bartneck2009measurement} and the human-robot interaction questionnaires~\cite{romero2015testing}. While the former has items focuses on observing the behavior of the robot (\textit{Anthropomorphism, Animacy, Likeability} and \textit{Perceived Intelligence}), the latter focuses on how the user interacts with the robot (\textit{Robot Motion, Conversation} and \textit{Sensation}). A shortened version, with relevant questions (\textit{Likeability, Perceived Intelligence, Conversation} and \textit{Sensation}) for a human coach, is used as the session experience questionnaire for \ac{HC} sessions. Questionnaires for both \acs{RC} and \ac{HC} sessions also measure how the participants felt in the beginning and end of a session with items: \textit{Anxious -- Relaxed} and \textit{Agitated -- Calm} to measure the effect of each session on them. The human coach filled the NASA TLX questionnaire~\cite{hart1988development} after each session recording the workload experienced during delivering the session both directly and while teleoprating robot.



\subsection{Robot and Teleoperation}
\label{teleop}
\begin{figure}[h!]
\centering
\includegraphics[width=\textwidth]{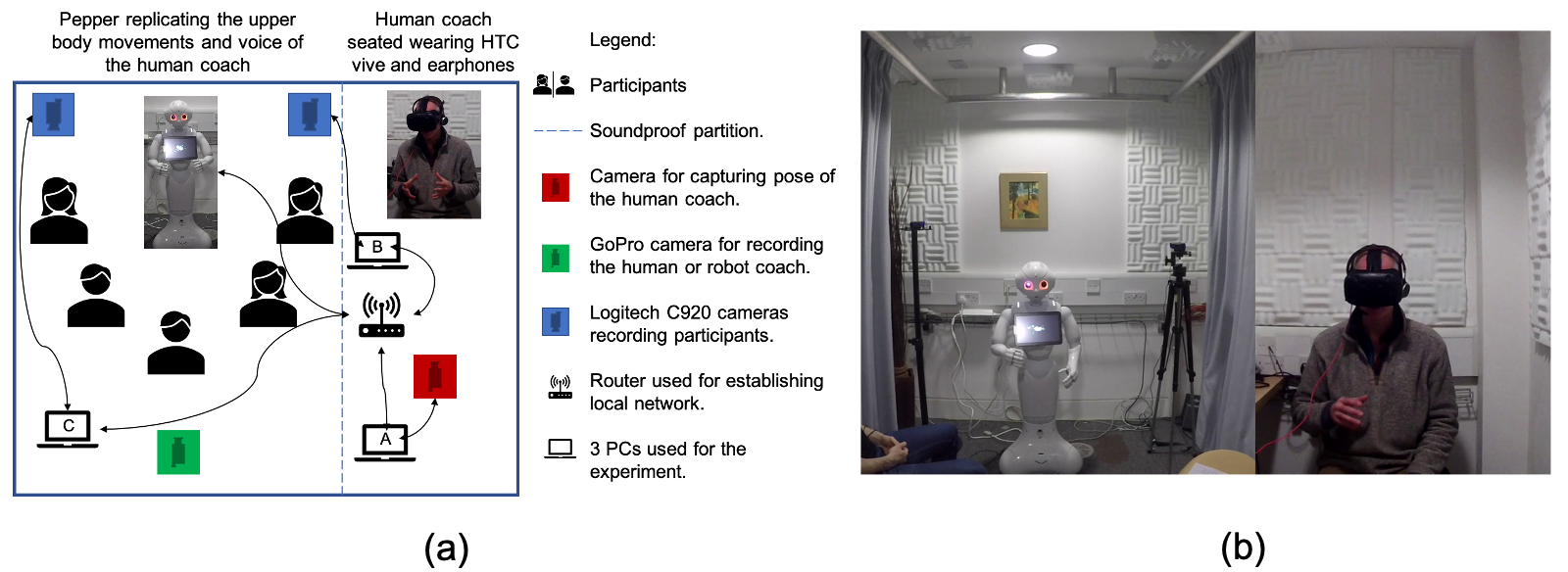}
\vspace{-4mm}
\caption{(a) Experiment setup during an \acs{RC} session with a maximum of $5$ participants. Pose, audio and video data are transmitted over a secure and encrypted network (using the router) for real-time processing. During \ac{HC} sessions, the human coach is seated directly in front of the participants and the room on the right side is not used.
(b) Side-by-side view of robot coach (L) and human coach (R) during teleoperation.} \label{expset}
\vspace{-5mm}
\end{figure}

The human coach teleoperated the Pepper robot to deliver \acs{RC} sessions. We implemented $3$ important components in the setup to enable real-time interactions between the teleoperating human coach and the participants. We established a secure, encrypted local network using a router to connect all the components needed for the transmission of video, audio and pose data between the robot and the human coach (see Fig.\ref{expset}). 

The human coach was seated in front of a camera used to capture and estimate his upper-body and head pose using the `Lifting from the deep' algorithm~\cite{tome2017lifting}. The algorithm initially worked for a single image was modified to estimate frame-by-frame pose in real-time. The estimated pose coordinates is interpolated from $3$ Hz to $30$ Hz to obtain smoother movements. Joint angles were then estimated based on \et{Ondras}~\cite{ondras2020audio} and projected onto the robot\footnote{\scriptsize{http://doc.aldebaran.com/2-5/naoqi/motion/almotion.html}}. Angle constraints were imposed to avoid dangerous or unnatural movements of the robot resulting from spurious detection. Additionally, the head pitch is hard-coded to $5$ limited angles; centre: [${-10^\circ}$,~${10^\circ}$] $\rightarrow~{0^\circ}$, two to the right of the robot: [${-10^\circ}$,~${-30^\circ}$]~$\rightarrow{-20^\circ}$ and [${\leq-30^\circ}$]~$\rightarrow~{-35^\circ}$, and two to the left of the robot: [${10^\circ}$,~${30^\circ}$]~$\rightarrow~{20^\circ}$ and [${\geq30^\circ}$]~$\rightarrow~{35^\circ}$, based the seats of the participants. The head yaw was fixed to ${0^\circ}$ to rectify the inaccuracies in estimating head pose by the algorithm. 
Real-time image frames were acquired from the camera on the forehead of the Pepper robot using the NAOqi vision API\footnote{\scriptsize{http://doc.aldebaran.com/2-4/naoqi/vision/alvideodevice.html}}, recording at $30$ FPS. The obtained frames are stacked horizontally with slight superposition (percentage of the pixels superposed estimated empirically) to adapt the robot's vision to be projected onto the HTC Vive VR headset\footnote{\scriptsize{https://www.vive.com/uk/product/\#vive\%20series}}. This allows the human coach teleoperating the robot to see `from the robot's eyes'. 

We established a SIP protocol-based audio call using  Jisti\footnote{\scriptsize{https://github.com/jitsi/jitsi-meet/blob/master/doc/sipgw-config.md}} over a local network between the teleoperator and the robot for the human coach to talk to the participants (through the robot) and listen to what they are saying. Due to bandwidth restrictions and audio feedback issues resulting from using the mic and speakers on the Pepper robot, we used an external speaker/mic placed behind the Pepper, hiding it from the participants, giving the illusion of the robot speaking with the participants.

\subsection{Data Analysis}
For analysing longitudinal changes in perceptions towards the human and robot coaches, we 
use the data collected from the Session Experience and the 20-item personality questionnaire filled by the participants along with the NASA TLX filled by the human coach for each session, each week.

We divided the questions in the \acs{RC} Session Experience questionnaire into the $8$ sections, namely, Anthropomorphism, Animacy, Likeability and Perceived Intelligence (constituting Impressions); Feelings in the beginning vs. the end;  Robot Motion, Conversation and Sensation (constituting Interactions). Similarly, we divided the questions in the \ac{HC} Session Experience questionnaire into four sections -- Impressions, Feelings in the beginning vs. End, and Interaction - Conversation and Interaction - Sensations. 
For further analysis, we used average scores for all items in each section. Impression and Interaction scores are combined together as \textit{Perception Scores}. We use the scores for Feelings in the beginning vs. the end to study the effect of mindfulness sessions on the participants. Using these evaluations, we investigate the following hypotheses:
\begin{itemize}[leftmargin=0.4cm]
\item H1: Perception scores of human coach do not vary significantly with time.
\item H2: Perception scores of robot coach change significantly with time.
\item H3: Mindfulness sessions promote relaxation and calm in both \ac{HC} and \acs{RC} sessions.
\item H4: Personality traits of the participants influence the perception scores of the robot coach.
\end{itemize}
\vspace{-6mm}

\subsubsection{Statistical Analysis}
To investigate H1 and H2, we used \acf{RCM} which was found to be very useful in analyzing longitudinal data with both inter- and intra- individual variability and missing data points. We assume that the data points are missing at random and therefore does not lead to bias in parameter estimates. We follow the model building approach illustrated by \et{Bliese}~\cite{bliese2002growth}. We present the results from random intercept model which models the individual variability in the overall levels of their perception ratings. The univariate random-intercept models of the perception scores are extended to test the interaction effect of the personality scores on the intercept variation, to study the effect on overall individual level and slope variation, to study the effect on growth patterns of the perception scores.
\vspace{-2mm}

\section{Results}
\begin{figure}[htb]
    \centering 
\begin{subfigure}{0.33\textwidth}
  \includegraphics[width=\linewidth]{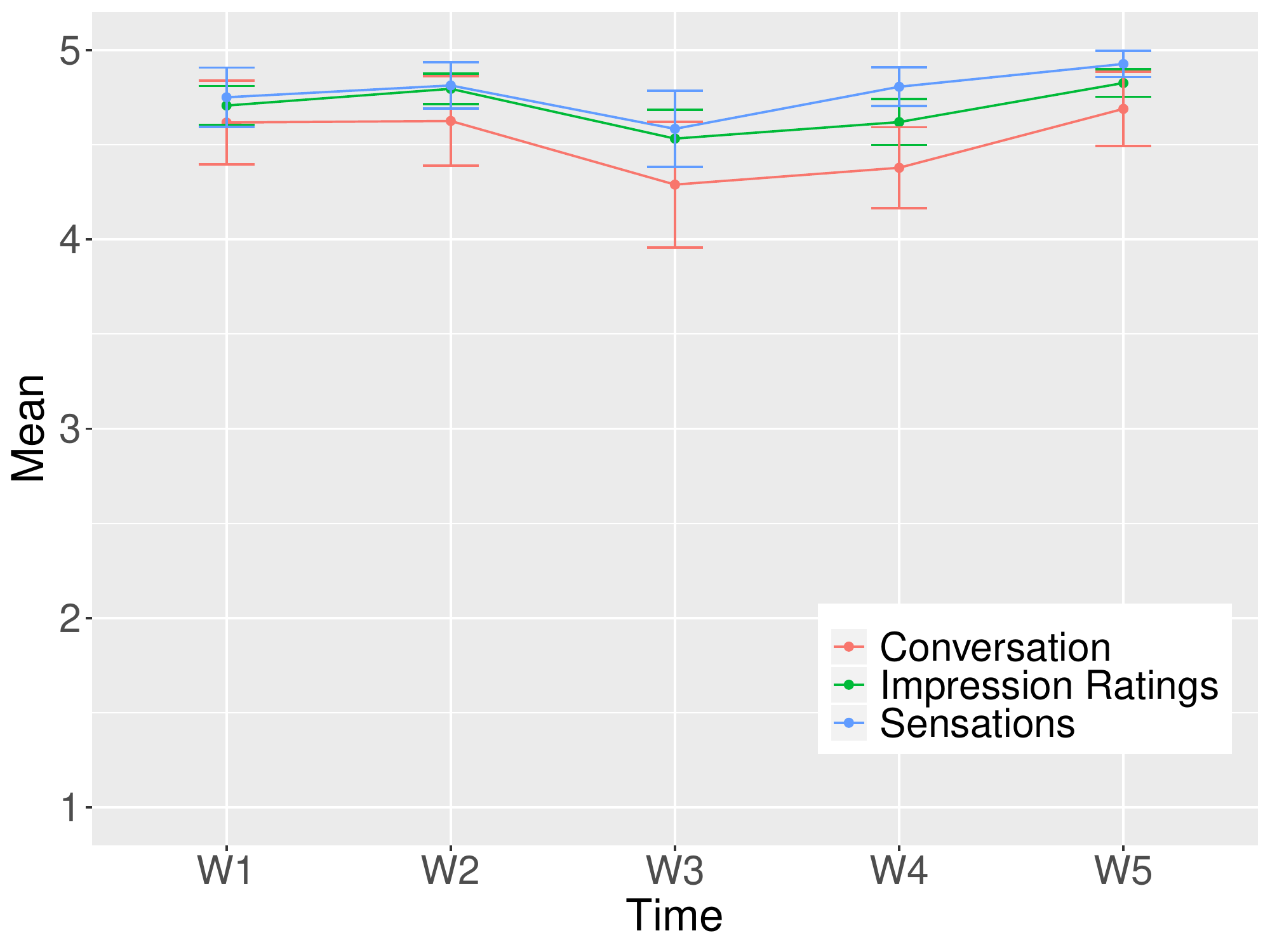}
  \caption{}
  \label{fig:2a}
\end{subfigure}\hfil 
\begin{subfigure}{0.33\textwidth}
  \includegraphics[width=\linewidth]{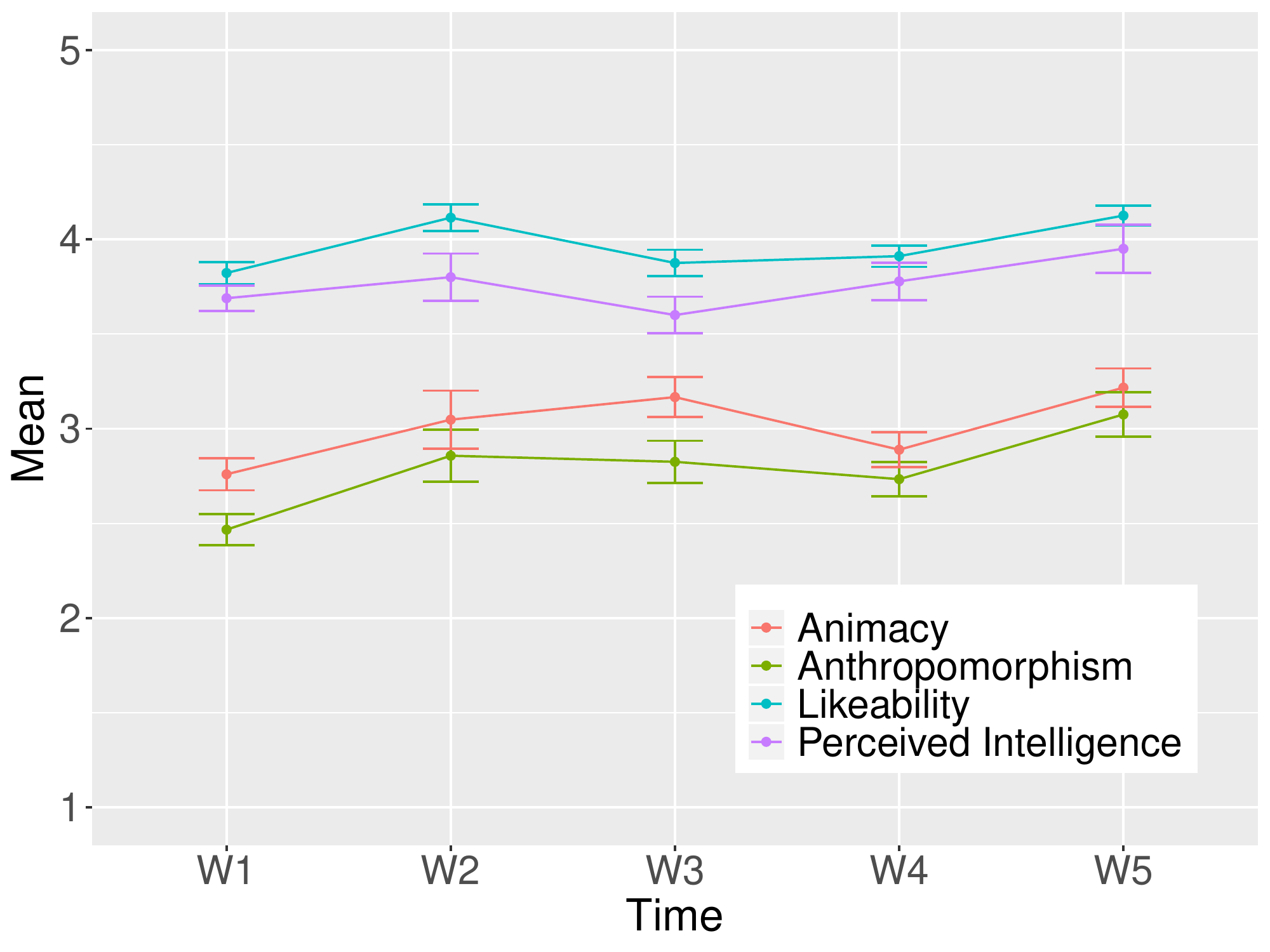}
  \caption{}
  \label{fig:2b}
\end{subfigure}\hfil 
\begin{subfigure}{0.33\textwidth}
  \includegraphics[width=\linewidth]{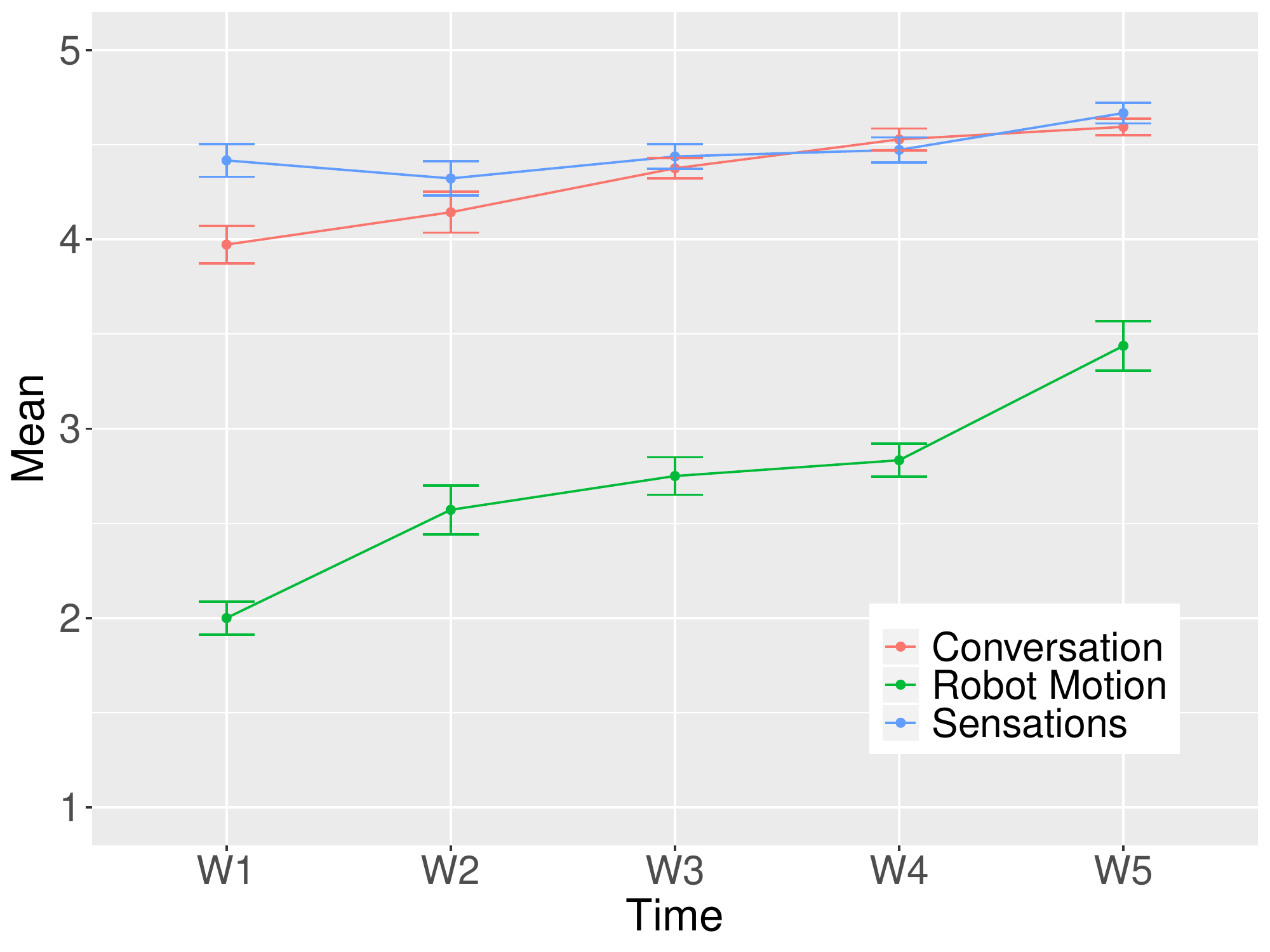}
  \caption{}
  \label{fig:2c}
\end{subfigure}

\medskip
\begin{subfigure}{0.33\textwidth}
  \includegraphics[width=\linewidth]{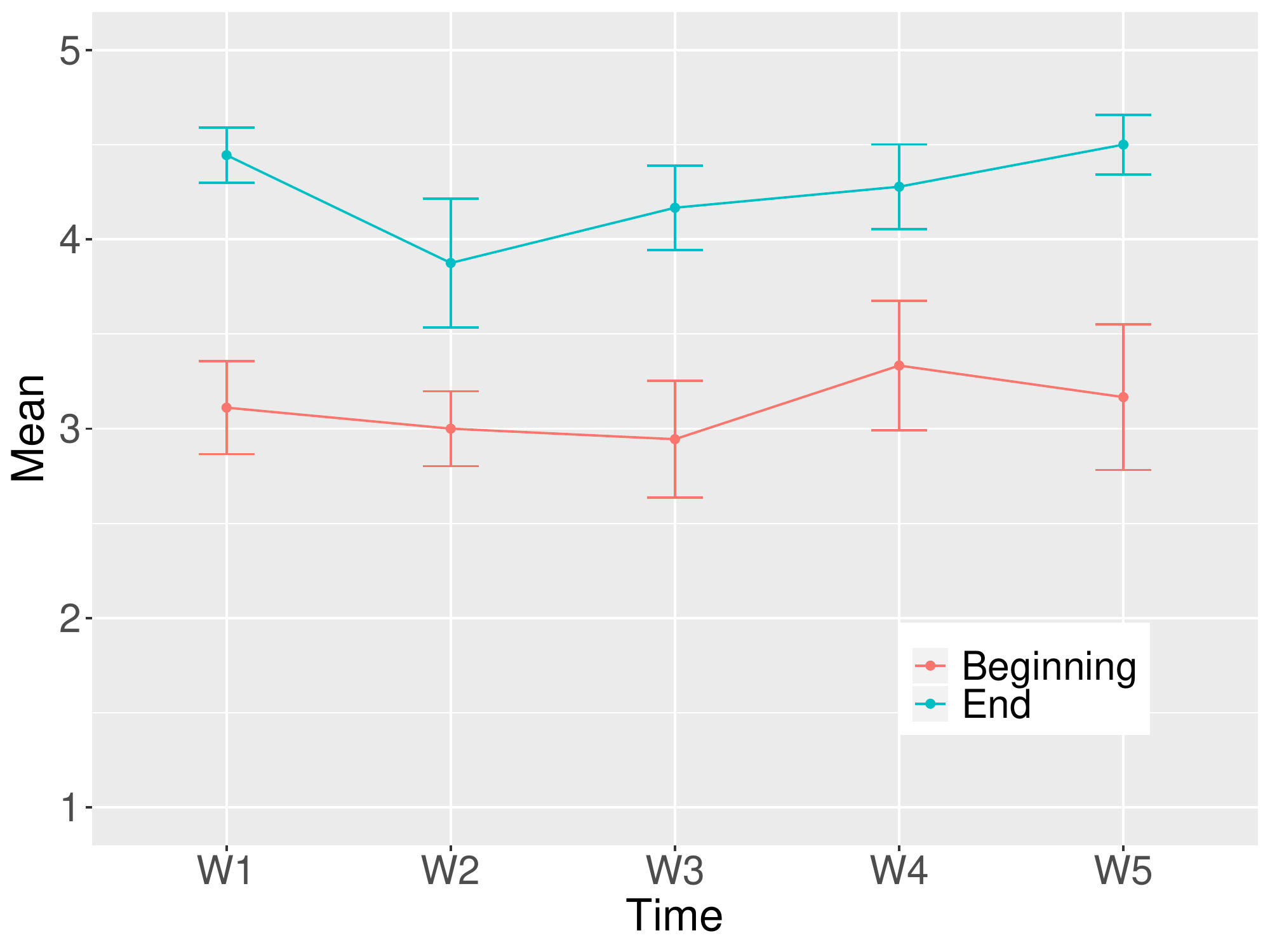}
  \caption{}
  \label{fig:2d}
\end{subfigure}\hfil 
\begin{subfigure}{0.33\textwidth}
  \includegraphics[width=\linewidth]{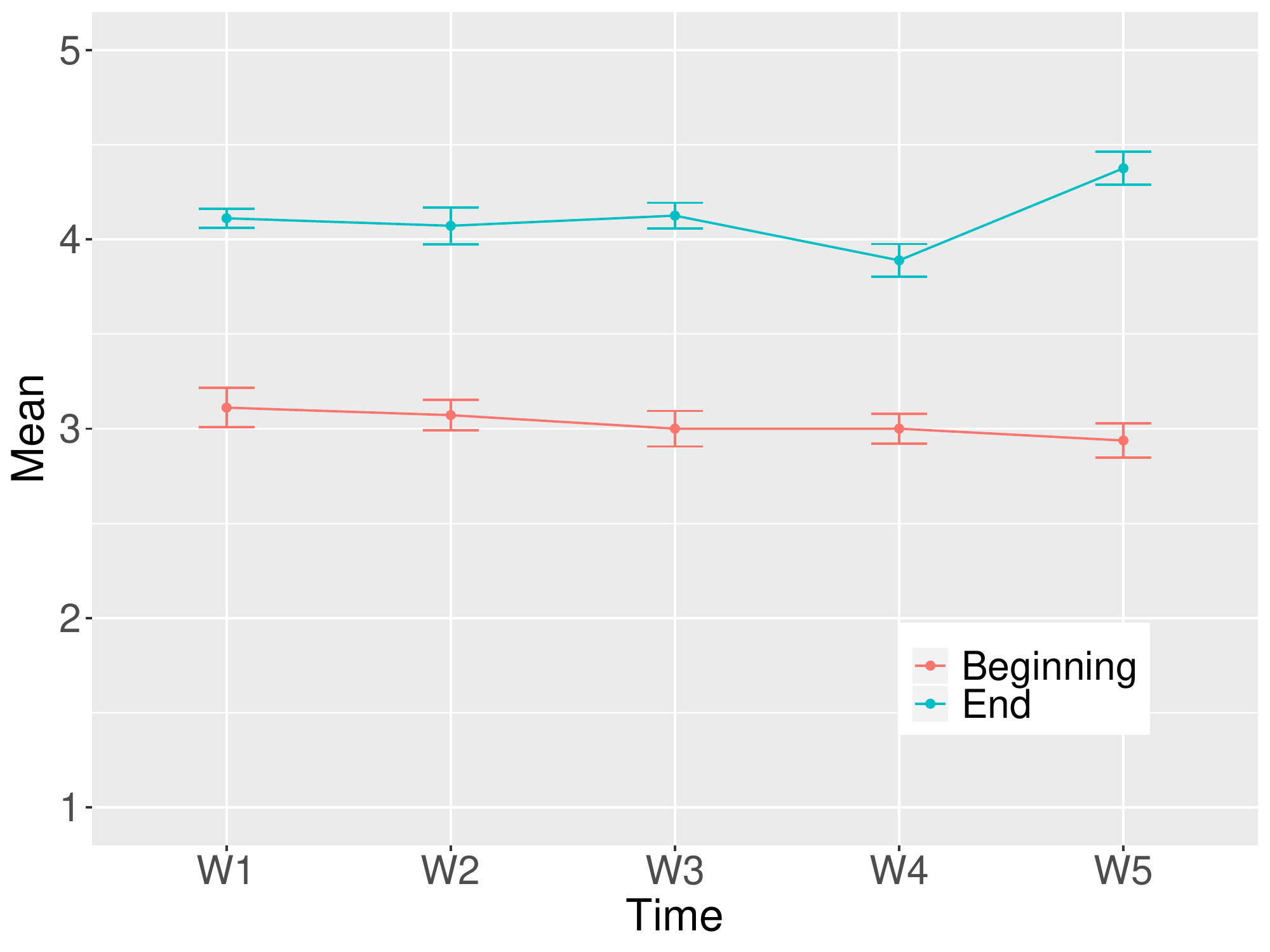}
  \caption{}
  \label{fig:2e}
\end{subfigure}\hfil 
\begin{subfigure}{0.33\textwidth}
  \includegraphics[width=\linewidth]{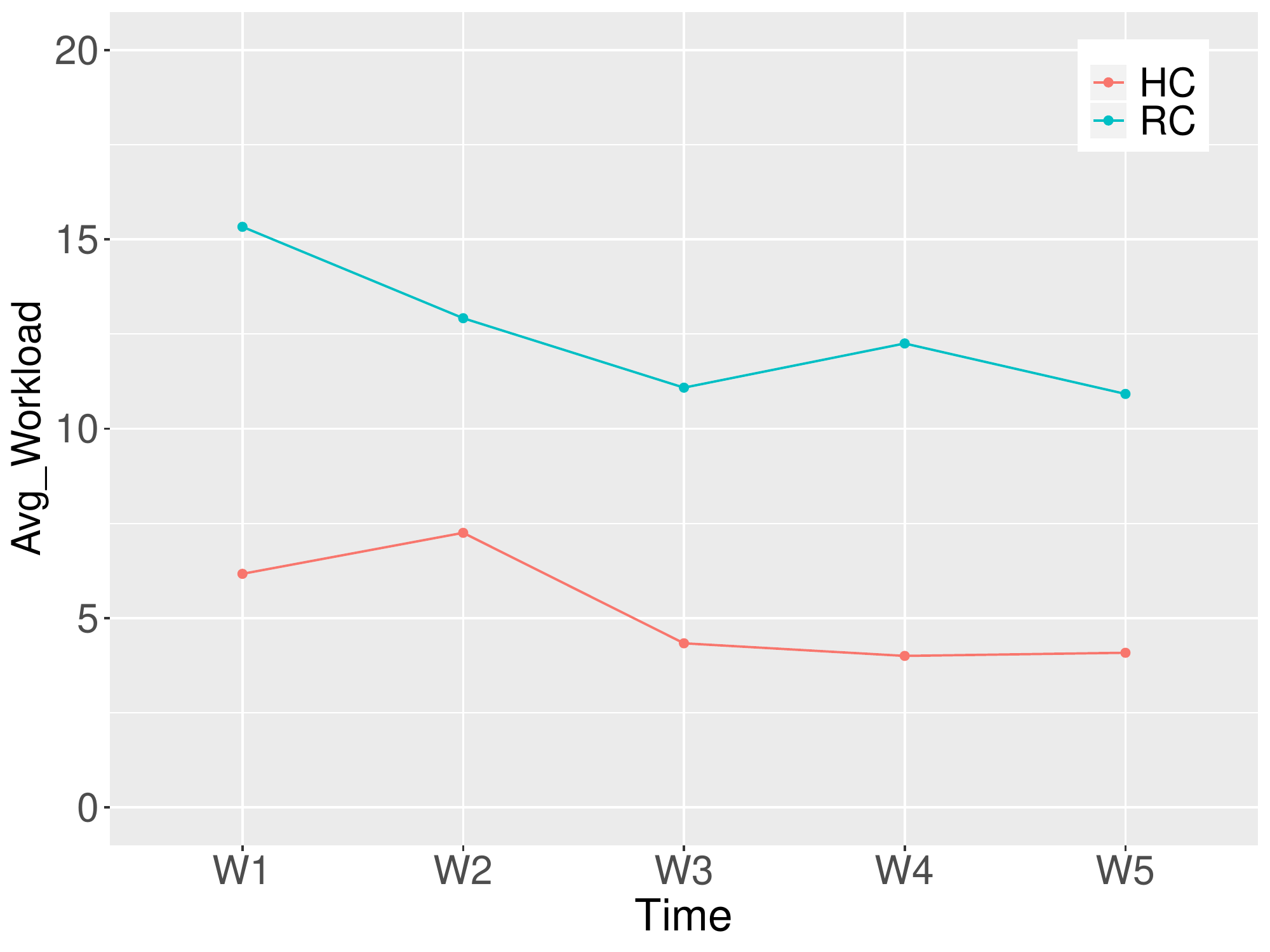}
  \caption{}
  \label{fig:2f}
\end{subfigure}
\caption{(a) Impression and Interaction scores for HC sessions. (b) Impression and (c) Interaction scores for RC sessions. Beginning vs. End feelings scores for (d) HC and (e) RC sessions. (f) Perceived Workload during HC and RC sessions. }
\label{fig:images}
\vspace{-3mm}
\end{figure}

\subsubsection{Interactions with the Human Coach (H1):}
The changes in the Impression scores, Interaction - Conversation and Sensation scores over time are shown in Fig. \ref{fig:2a}. No significant changes were witnessed in the Impression scores ($t=0.194; DF=34; p=0.84$), Interaction-Conversation score ($t=-0.39; DF=34; p=0.69$) and Interaction-Sensation scores ($t=1.28; DF=34; p=0.20$) of the human coach with time. Moreover, the impression and interaction scores stayed consistently high ($mean > 4.0$ for all the sessions on a scale of $1-5$). This suggests that the participants reacted consistently positively towards the human coach right from the first session.
\vspace{-2mm}

\subsubsection{Interactions with the Robot Coach (H2):}

\label{robot_results}
The longitudinal changes in the Impression score in terms of Anthropomorphism, Animacy, Likeability and Perceived Intelligence are shown in Fig. \ref{fig:2b}. Qualitatively, we see that the participants rated the robot low on Anthropomorphism and Animacy but high on Likeability and Perceived intelligence. This suggests that our design for the robotic coach invoked positive responses from the participants despite feeling robot-like. Further, we did not see any significant changes in the Impression scores with time: Anthropomorphism ($t=1.68; DF=31; p=0.10$), Animacy ($t=1.23; DF=31; p=0.23$), Likeability ($t=0.82; DF=31; p=0.42$) and Perceived Intelligence ($t=1.18; DF=31; p=0.25$). This is not surprising as the appearance and functionality of the robot does not change across the 5 sessions.


The longitudinal changes in the Interaction scores, that is, Robot Motion, Conversation and Sensations are shown in Fig. \ref{fig:2c}. We found a significant effect for Robot Motion ($t=4.71; DF=31; p<0.05$) and Conversation ($t=2.57; DF=31; p<0.05$) with both the scores increasing over time. This suggests that with time, participants found that the robot moved and interacted more consistently as well as easier to converse with. 

\vspace{-2mm}

\subsubsection{Mindfulness sessions to promote relaxation and calm (H3):}
We used the average ratings of two items, namely, Anxious-Relaxed and Agitated-Calm scores, at the beginning and end of each session. We conducted a two-way repeated measures ANOVA with one factor as the Time (5 weeks) and the other factor as Instance (beginning vs. end). We found a significant effect of Instance, in that the scores at the end of the session were significantly higher than at the beginning for both \ac{HC} $(F=9.365; p=0.018)$ (see Fig. \ref{fig:2d}) and \acs{RC} groups $(F=24.64; p=0.008)$ (see Fig. \ref{fig:2e}) suggesting that the sessions made the participants more relaxed and calm. There is no observed effect over time.

\vspace{-2mm}

\subsubsection{Interaction between Perception and Personality Scores:}
To investigate the effect of participants' personality on their perception scores, we conducted a separate analysis by including each of the personality traits (extroversion, agreeableness, conscientiousness, neuroticism and openness), into the previous random intercept model of each attribute of the perception scores. We modeled the effect of personality traits on both the intercept and slope of the perception scores. Out of the five personality traits, Conscientiousness and Neuroticism showed significant effects on the perception scores. We found a significant negative effect of Conscientiousness on the intercept variability of Robot Motion ($t=-2.618; DF=7; p=0.032$). This suggests that people with higher conscientiousness scores gave low overall ratings for robot motion. However, no significant effect was found with respect to the changes across time.  Additionally, we found a significant negative effect of Neuroticism on the intercept ($t=-2.789; DF=7; p=0.027$) and slope ($t=-2.968; DF=7; p=0.021$) variability of Sensation and a significant positive interaction effect between Neuroticism and time on the slope variability of Robot Motion ($t=2.17; DF=30; p=0.038$). Neuroticism or low emotional stability negatively effects on the overall session experience and also effects how the perception score for robot motion changes with time.

\vspace{-2mm}

\subsubsection{Qualitative workload comparison:}

The NASA TLX scores collected from the human coach in both \ac{HC} and \acs{RC}, indicate an increased perceived workload for the \acs{RC} sessions (see Fig.~\ref{fig:2f}). Although, no statistical analysis was conducted on this data obtained from only one coach, our discussions with the coach highlight this increase as a direct consequence of having to adapt to the use of additional equipment (such as the VR headset) in the \acs{RC} sessions.


\section{Conclusions and Future Directions}

In this paper, we report on research work that bridges two important enablers for improving wellbeing: telepresence robotics and mindfulness coaching. We follow an iterative design process, with the ultimate goal of creating a fully autonomous and adaptive mindfulness robot coach. The first version of the robot coach was implemented using teleoperation with multimodal interaction capabilities (voice and head/arm gestures) that enables us to test our initial hypotheses and better understand the context as well as the expectations and the challenges faced.

Our results show that the impression scores corresponding to items adapted from the Godspeed questionnaire did not change with time for both the human and the teleoperated robot coach. These ratings correspond to characteristic attributes of the coaches and therefore the perceptions regarding these attributes do not change for similar type of interactions across time. Godspeed ratings have usually been used to compare different robots or the same robot performing different tasks~\cite{weiss2015meta} and might not be suitable for evaluation of long-term HRI. 

The robot coach is rated high for Likeability, showing characteristics such as pleasantness, voice-only communication, synchronous movements which are rated highly likeable~\cite{chatterji2019effectiveness}; and Perceived Intelligence, as it is being teleoperated by the human coach~\cite{weiss2015meta}. However, Anthropomorphism and Animacy ratings were low for the robot coach. In our future work, we would like to focus on customized generation of gestures where there is a congruence between the speech and movements of the robot to enhance natural perception. For example, \et{Ondras}~\cite{ondras2020audio} demonstrated that viewers preferred machine-like movements generated to match machine-like synthetic speech of the robot. 

We also found a longitudinal increase in the interaction ratings for Robot Motion and Conversation. We did not observe the novelty effect, that is, a drop in ratings of the robot over time. Our post-study interviews with the participants from the \acs{RC} sessions revealed a preference for the robot that could move synchronously while speaking and could maintain eye contact during conversations. The participants appreciated that they can maintain a coherent conversation with the robot where it checked on them periodically and they could give feedback about how they were doing. These results suggest that participants accepted the use of a robot coach to deliver mindfulness sessions over time but expect functionalities such as generation of natural language and expressive gestures for interaction from an autonomous platform~\cite{foster2019natural,van2018generic}. 
We also found significant links between participants' personality traits and their perception scores which suggests that person-specific customization should be included in the design of an autonomous coach. 

The human coach is rated consistently high in terms of both impressions and interactions. We also did not observe any novelty effect, that is, drop in ratings with time towards the human coach. \et{Van Alderen}~\cite{van2014role} demonstrated that characteristics such as the embodiment of a non-judgemental and compassionate stance as well as empowerment and non-reactivity play an important role in the acceptance of a mindfulness teacher. The mindfulness sessions in our study are delivered by a certified and experienced coach who has personally been practising mindfulness for many years. The sessions were therefore reported to be beneficial to the participants in both the groups.We note that care should be taken while moving towards designing an autonomous platform, where translating the experience of the human coach into robot delivered sessions is only possible if the human coach has supervisory control over the robot's delivery. 
The human coach in our experiments reported that although teleoperating gets easier with time, it is still tiresome compared to delivering the session directly. These suggestions motivate us to investigate strategies for creating platforms with supervised autonomy~\cite{esteban2017build,cao2019robot}.

\subsection*{Acknowledgements}
I.P Bodala and H. Gunes are supported by EPSRC under grant EP/R030782/1. N. Churamani is supported by EPSRC under grant EP/R513180/1 (2107412). The authors also gratefully acknowledge support from Simon McKibbin for delivering the mindfulness sessions; Sinan Kalkan and Janko Ondras for advice on LFTD and joint angle estimation; Paul Bremner for advice on the audio setup; and Miriam Koschate-Reis and Elahe Naserianhanzaei for advice on longitudinal data analysis methods.

\begin{acronym}
\acro{ANOVA}{Analysis Of Variance}
\acro{ASD}{Autism Spectrum Disorder}
\acro{GLM}{Generalised Linear Model}
\acro{HC}{Human Coach}
\acro{RC}{Robot Coach}
\acro{RCM}{Random Coefficients Modelling}
\acro{SAR}{Socially Assistive Robots}
\end{acronym}

\bibliographystyle{splncs04}
\bibliography{main.bib}

\end{document}